\documentclass [aps,prx,twocolumn,groupeauthor,showpacs,superscriptaddress,longbibliography]{revtex4-1}

\usepackage{amsmath}
\usepackage{amssymb}
\usepackage{mathtools}
\usepackage{latexsym}
\usepackage{bm}

\usepackage{graphicx}

\DeclareMathOperator{\Tr}{Tr}
\DeclareMathOperator{\bbeta}{\boldsymbol\beta}
\DeclareMathOperator{\aalpha}{\boldsymbol\alpha}

\begin{document}
\title{Bypassing the energy-time uncertainty in time-resolved photoemission}
\author{Francesco Randi}
\affiliation{Department of Physics, Universit\`{a} degli Studi di Trieste, 34127 Trieste, Italy}
\affiliation{Max Planck Institute for the Structure and Dynamics of Matter, 22761 Hamburg, Germany}
\author{Daniele Fausti}
\affiliation{Department of Physics, Universit\`{a} degli Studi di Trieste, 34127 Trieste, Italy}
\affiliation{Sincrotrone Trieste SCpA, 34127 Basovizza, Italy}
\author{Martin Eckstein}
\affiliation{Max Planck Institute for the Structure and Dynamics of Matter, 22761 Hamburg, Germany}
\affiliation{University of Hamburg-CFEL, 22761 Hamburg, Germany}
\email{martin.eckstein@mpsd.cfel.de}


\begin{abstract}
The energy-time uncertainty is an intrinsic limit for time-resolved experiments imposing a tradeoff between the duration of the light pulses used in experiments and their frequency content. In standard time-resolved photoemission this limitation maps directly onto a tradeoff between the time resolution of the experiment and the energy resolution that can be achieved on the electronic spectral function. 
Here we propose a protocol to disentangle the energy and time resolutions in photoemission. We demonstrate that dynamical information on all timescales can be retrieved from time-resolved photoemission experiments using suitably shaped light pulses of quantum or classical nature.  
As a paradigmatic example, we study the dynamical buildup of the Kondo peak, a narrow feature in the electronic response function arising from the screening of a magnetic impurity by the conduction electrons. After a quench, the electronic screening builds up on timescales shorter than the inverse width of the Kondo peak and we demonstrate that the proposed experimental scheme could be used to measure the intrinsic timescales of such electronic screening. The proposed approach provides an experimental framework to access the nonequilibrium response of collective electronic properties beyond the spectral uncertainty limit and will enable the direct measurement of phenomena such as excited Higgs modes and, possibly, the retarded interactions in superconducting systems.

\end{abstract}
\date{\today}
\maketitle

\section{Introduction}
The electronic response function at low energies is a standard tool to address the physical mechanism leading to the exotic macroscopic properties observed in complex materials. The onset of superconductivity, charge density wave or giant magnetoresistance, which arise from both the electron-electron interactions as well as from the interactions between electrons and other degrees of freedom, is often reflected in anomalous features in the low energy electronic response function~\cite{Damascelli2003,Avella}. In those systems, the population dynamics and the electronic gaps are often not ``factorizable'', i.e. the excited state population dynamically renormalizes the electronic gaps~\cite{Giannetti2016}. This aspect makes time-resolved photoemission particularly appealing among non-equilibrium techniques, as it allows to simultaneously measure the time-dependent band structure and the evolution of the non-equilibrium distribution of electronic excitations. Various time- and angle-resolved photo-emission experiments have revealed the collapse of charge density waves and Mott gaps~\cite{Petersen2011, Schmitt2008, Rohwer2011} or the reconstruction of the gap at photo-induced metal-insulator transitions~\cite{Perfetti2006,Wegkamp2014}. 

Time domain pump and probe photo-emission spectroscopy suffers from the intrinsic limitation imposed by energy-time uncertainty relation. In fact, although nowadays we have technologies to produce light pulses shorter than 100 attosecond~\cite{Chini2014}, the shorter the pulses are, the larger is their bandwidth. In standard time domain photoemission experiments the large energy content of the probing pulse maps directly onto a large spread in the energy content of the photo-emitted electron. This results in a poor energy resolution that averages out the spectroscopic feature associated to low energy gaps.

For this reason up to now, time domain studies have been limited to study 
the dynamics of spectral features
which are slower than the inverse of their characteristic frequency. In general, one could say that the time-evolution takes place {\em beyond the spectral uncertainty limit} if relevant degrees of freedom evolve faster than the inverse width of their spectroscopic fingerprint.  

A paradigmatic example of such situation is the Kondo effect, i.e., the screening of an impurity magnetic moment, which is reflected only in a narrow spectral feature close to the Fermi energy: Since a pulse of duration $\Delta t$ has a minimum bandwidth  $\Delta \omega = 1 / \Delta t$, it is generally believed that the dynamics of the Kondo screening can be measured only on timescales slower than the inverse of its spectral width, while it is known that the screening itself can evolve on a shorter timescale~\cite{Lechtenberg2015}. A similar issue can arise in the case of the condensate in standard BCS superconductors, whose dynamics is believed to be observable only on timescales which are slower than the inverse of the gap frequency, $\hbar/\Delta_\text{sc}$. On the other hand it is expected that gap amplitude oscillations at the gap frequency should follow a sudden quench of the superconducting gap~\cite{Matsunaga2013, Matsunaga2014}.

In this manuscript we show that the  
tradeoff between the resolution on the photoelectron energy and the temporal resolution is a characteristic of standard pump-probe photoemission performed with pulses with Gaussian envelopes,
and that it can be circumvented using properly shaped light pulses. 
Here, we extend the well established concepts of multidimensional optical spectroscopy~\cite{MukamelBook,Mukamel1999} to time-resolved photoemission experiments.
The idea leading this theoretical proposal is that the full information about the single particle dynamics in the solid, contained in the Green's function $G(t,t')$, can be accurately characterized entirely in the time domain and the limitations of the mixed time-frequency approach used in standard time domain experiments can be overcome. In particular, we show that a real-time measurement of the Green's function can be experimentally implemented using the interference photo-emission signal from a double probe pulse, obtained by splitting, delaying and recombining a pulse of arbitrarily short duration. 

We illustrate this double probe pulse photo-emission scheme addressing the buildup time of the Kondo peak and highlight how this scheme could have a more general relevance for experiments addressing the true timescales of destruction of electronic order in Mott insulators such as 1$T$-TaS$_2$ dichalcogenide~\cite{Sohrt2014} and superconductors~\cite{Matsunaga2013,Matsunaga2014,Kemper2015}.

In addition to the experiments exploiting the modulation of the classical intensity we envision experiments taking advantages of light pulses with 
statistical or 
quantum correlations. Exploiting quantum correlations of light both in controlling and measuring the properties of matter has a wide application in other fields~\cite{Haroche2012,Wineland2012,Schnabel2010,Treps2002,Esposito2015}. Our proposal represents the first attempt to exploit 
non-coherent states of
light to address specific spectroscopic features in time-resolved photoemission experiments.

%%%%%%%%%%%%%%%%%%%%%%%%%%%%%%%%%%%%%%%%%%%%%%%%%%%%%%
\section{Generalized time-resolved photoemission}
\label{sec:photoemission}
%%%%%%%%%%%%%%%%%%%%%%%%%%%%%%%%%%%%%%%%%%%%%%%%%%%%%%

\subsection{Theoretical formulation} 

In a time-resolved angle-integrated photoemission experiment, one measures the probability $I(E,t_p)$ that an electron is emitted under the action of a short probe pulse, as a function of the photoelectron energy $E$ and the time-delay $t_p$ between the probe pulse and a given excitation (e.g. the pump pulse). 
The signal $I(E,t_p)$ can be obtained using time-dependent perturbation theory in the light-matter coupling  \cite{FreericksKrishnamurthyPruschke2009, Eckstein2008c}.  If $s(t)e^{i\Omega t} + s^*(t)e^{-i\Omega t}$ denotes the time-profile of the probe vector potential with envelope $s(t)$ and centre frequency $\Omega$, one obtains (see Ref.~\cite{FreericksKrishnamurthyPruschke2009} and appendix \ref{appendix:derivation})
\begin{align}
I(E,t_p) 
\propto \!-i \!\!
\int \!\!dt\,dt'\,\,
e^{iE(t-t')} 
G
^<
(t_p\!+\!t,t_p\!+\!t')\,\,S(t,t').
\label{eq:photoemission1}
\end{align}
Here the kinetic energy $E$ is defined with respect to the energy $\hbar\Omega-W$ given by the frequency  $\Omega$ and the work function $W$, and 
$S(t,t')=s(t)s(t')^*$ is an autocorrelation function of the probe pulse, which acts as a {\em filter} determining how $e^{iE(t-t')}G^<(t,t')$ is sampled over the $(t,t')$-plane in the above integration.  
Furthermore,  $G^<(t,t') = i \langle f^\dagger(t')f(t) \rangle$ is the Green's function of the sample alone, where $f$ ($f^\dagger$)  is the annihilation (creation) operator for an electron in a given orbital of the system, $\langle \cdots\rangle $ denotes the expectation value in the initial state (at $t\to -\infty$), and the time evolution includes all nonequilibrium perturbations besides the probe, such as external pump laser fields. 
Equation \eqref{eq:photoemission1} can easily be extended by adding a sum over orbitals and matrix elements, but the latter are static and do not alter the following general discussion of the relation between the time-dependent electronic properties and the photoemission signal. Similarly, by inserting suitable matrix elements the results discussed in this work can be reformulated for angle-resolved photoemission~\cite{Note0}.

\begin{figure}
\includegraphics[width=0.35\textwidth]{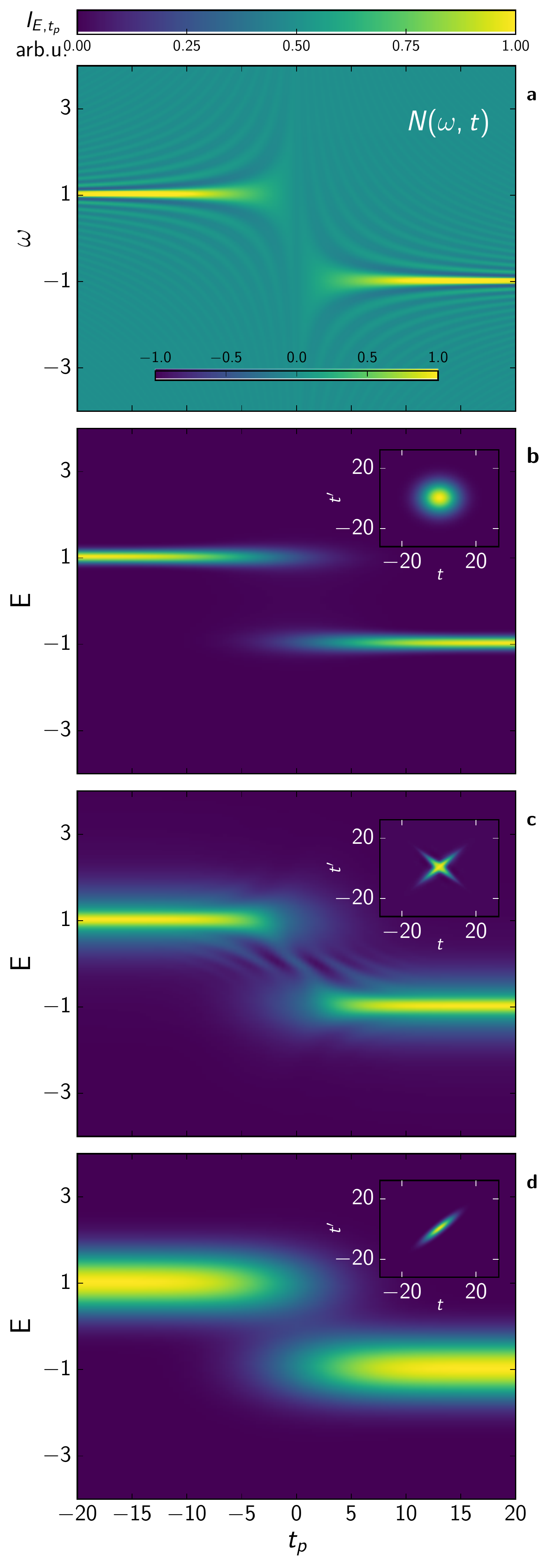}
\caption{
Illustration of the use of different filters $S(t,t')$ (pulses) in a photoemission experiment: 
The system of interest is a single level with
$H(t) = \text{sign}(t) \epsilon_0  c^\dagger c $,
which is initially occupied and then suddenly shifted in energy $(\epsilon_0=1)$. Panel \textbf{a)} shows the Wigner transform $N(\omega, t)$ of the Green's function of the system.
The remaining panels show time-resolved angle-integrated photoemission spectra obtained with different pulses ($S(t,t')$ is shown in the inset), as a function of photoelectron energy and time. 
\textbf{b)} Classical Gaussian probe pulse. The observed switching is resolution limited.
\textbf{c)} Non-separable positive definite filter S(t,t'), which, thanks to its off diagonal structure, produces some of the features of the $N(\omega, t)$ in the photoemission spectrum.  
\textbf{d)} Incoherent pulse with the same intensity profile in time as the coherent pulse in \textbf{b)}, which yields a lower energy resolution.
}
\label{fig:quantum}
\end{figure}

If the probe pulse is modelled by a Gaussian $s(t) =\exp(-t^2/2\Delta t^2)$ with duration $\Delta t$, Eq.~\eqref{eq:photoemission1} can be transformed into a mixed time frequency representation,
\begin{align}
\label{eq:phototw}
I(E,t_p) \propto \int d\omega\, dt\,\, 
N(E+\omega,t_p+t)\,\,
e^{-\frac{t^2}{\Delta t^2}} e^{-\omega^2\Delta t^2} ,
\end{align}
where $N(\omega,t) = \int \frac{ds}{2\pi i} \,e^{i\omega s} \,G^<(t+s/2,t-s/2)$ is the Wigner transform of the Green's function, which could be referred to as a time-dependent occupied density of states. (At equilibrium, $N(\omega)=A(\omega) f(\omega)$ is the product of the spectral function $A(\omega)$ and the Fermi distribution function.) Equation~\eqref{eq:phototw} emphasizes the origin of the uncertainty limit in time-resolved photoemission: The signal $I$ is related to the underlying spectral information $N(\omega,t)$ by a filter which is subject to the uncertainty $\Delta\omega=1/\Delta t$ in ($t,\omega$) plane. This is illustrated in Fig.~\ref{fig:quantum}, for the simple example of an occupied level which is suddenly shifted in binding 
energy from $\epsilon_i$ to $\epsilon_f$. For simplicity, we choose the zero of enegy such that $\epsilon_i = +1$  $(< \epsilon_{\mathrm{vacuum}})$ and $\epsilon_f = -1$. Figure \ref{fig:quantum}a and b show the Wigner transform $N(\omega, t)$ and  the result of time-resolved photoemission with a Gaussian pulse, respectively.  While $N(\omega, t)$ contains the full information about single-particle properties in the time-evolving quantum state, this information can no longer be easily reconstructed from the photoemission intensity. However, one may now ask whether probe pulses can be appropriately devised to shape $S(t,t')$ in order to access the underlying information in $G^<(t,t')$ or  $N(\omega,t)$ in a more flexible way. There is clearly no physical pulse which could yield $N(\omega, t)$ in a single intensity measurement, since $N(\omega, t)$ can take negative values,  while $I(E,t)$ is a nonnegative probability~\cite{Note1,Eberly1977,Mukamel1996,Dorfman2012}. Nevertheless, in the following sections we will show that there are filters which may be more useful than the Gaussian one in Eq.~\eqref{eq:phototw} in analyzing ultrafast dynamics, since they allow to extract different kind of information, or can be used to ``tomographically'' reconstruct $G^<(t,t')$ and  equivalently $N(\omega, t)$ with a series of measurements.

\subsection{Double-probe photoemission: Tomography of $G^<(t,t')$}
\label{sec:tomography}

\begin{figure}
\includegraphics[width=0.35\textwidth]{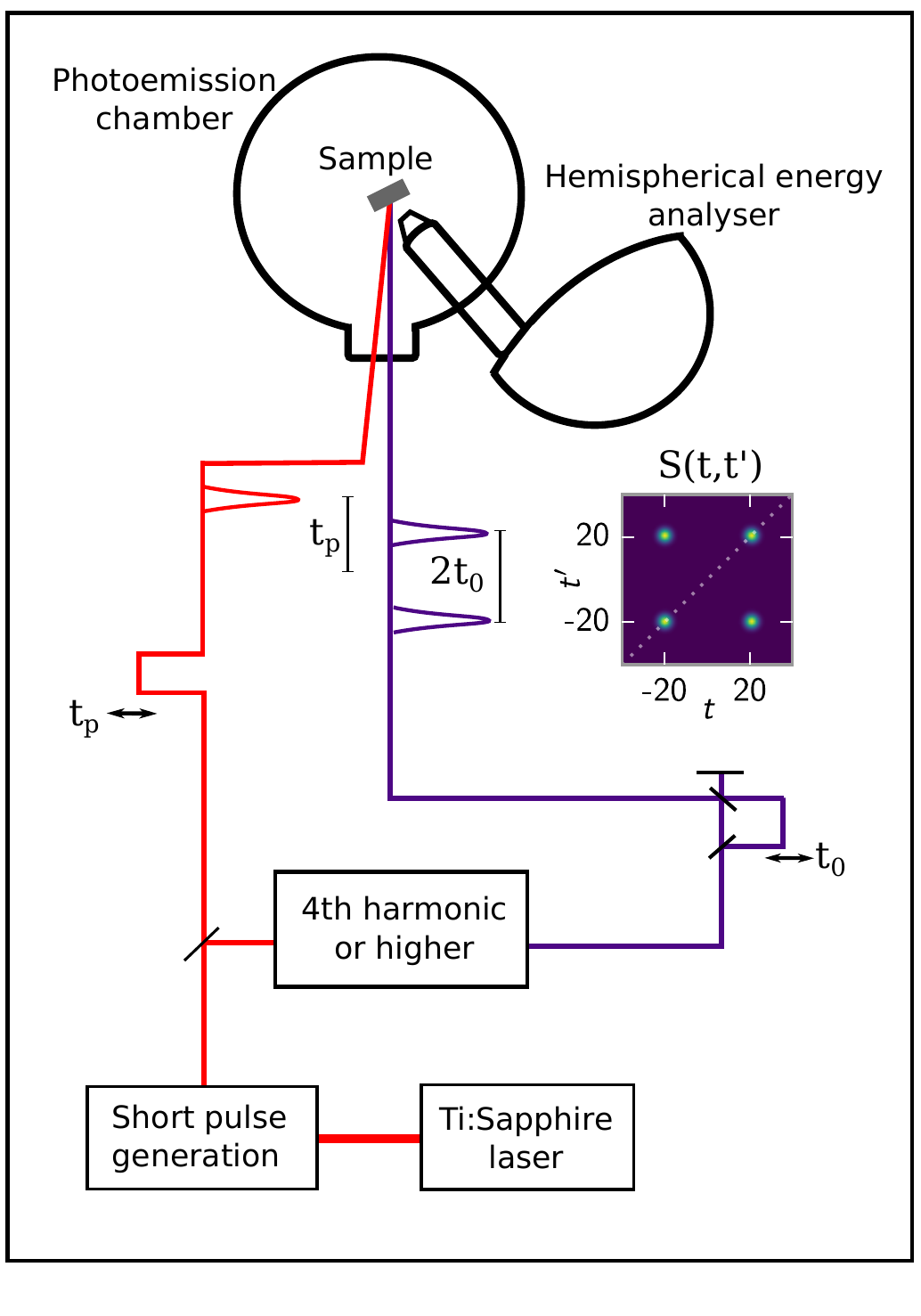}
\caption{
Schematic sketch of a Ti:Sapphire laser based experimental set-up that uses the double probe pulses to perform the tomography of the Green's function, by varying $t_p$ and $t_0$.
}
\label{fig:setup}
\end{figure}

In this section we propose a probing scheme that allows to determine $G^<(t,t')$ in real time, based on a set of measurements with two probe pulses, which are separated in time, but have a fixed phase relation. The procedure is realistically implementable from the experimental point of view with the current technology, using splitted, delayed and recombined pulses, as shown schematically in Fig.~\ref{fig:setup}. 
It resembles what is usually done in the established field of multidimensional optical spectroscopy, albeit the fact that what is measured is the number of emitted electrons.
If $s_0(t)$ denotes the envelope of the single pulse, the total envelope $s(t)$ in Eq.~\eqref{eq:photoemission1} is given by
\begin{equation}
\label{eq:double_pulse}
s(t) \propto 
s_0(t-t_0) +  e^{i\varphi} s_0(t+t_0)
\equiv
s_+(t) +  e^{i\varphi} s_-(t)
\end{equation}
where $\varphi$ is 
the relative difference of carrier envelope phase between the two probe pulses,
and  $2t_0$ is the temporal separation.
Using Eq.~\eqref{eq:photoemission1}, the photoemission intensity obtained with this double pulse is given by the sum of the photoemission $I_{\pm}(E,t_p)$ obtained with the individual pulses $s_{\pm}(t)$ as if they were used separately, and an {\em interference signal}
\begin{align}
I_\text{intf}(E,t_p) 
\propto
\text{Im} \,
\Big[
e^{iE2t_0-i\varphi}
\int \!\!dtdt' s_0(t)s_0(t')^* \times
\nonumber \\
e^{iE(t-t')} 
G^<\big((t_p+t_0)+t,(t_p-t_0)+t'\big)
\Big].
\label{eq:photoemission-12}
\end{align}
This is determined by the particular shape of the filter S(t,t') (inset of Fig.~\ref{fig:setup}): 
The first contribution samples $G^<(t,t')$ on the $t=t'$ diagonal in $(t,t')$-space, while the interference contribution samples it out of the diagonal.
Note that $\varphi$ is the relative difference of carrier envelope phase between the two probe pulses, and not an absolute carrier envelope phase. Therefore, if the phase relation between the two probes is fixed, $I_{\mathrm{intf}}$ does not vanish upon averaging over many laser shots. As an example, if the probe photon energy is 6 eV ($\lambda_{\mathrm{vac}}\simeq$ 200 nm), the stability of the mirrors of the delay stage in Fig.~\ref{fig:setup} controlling $t_0$ should be of approximately 100 nm.

The above described
result is understood most easily in the limit of extremely short pulses  $s_0(t)=\delta(t)$.  A single  ultrashort pulse retrieves only the time-dependent density $n_f=\langle f^\dagger f \rangle$ [c.f.~Eq.~\eqref{eq:photoemission1}], so that $I_{\pm}(t_p,E) \propto n_f(t_p\pm t_0)$, while 
\begin{align}
\label{eq:pestomo}
I_\text{intf}(E,t_p) \propto 2 \, \text {Im}\Big[\, G^<(t_p+t_0,t_p-t_0) e^{i2Et_0 - i\varphi}\Big].
\end{align}
The interference contribution can be extracted in two ways. On the one hand, it can be identified by its oscillating dependence on the photoelectron final state energy. On the other hand, the diagonal terms $I_{\pm}$ can be obtained from an independent measurement with a single probe pulse experiment. Even without any knowledge of the relative phase $\varphi$, one can thus obtain the absolute value $|G^<(t_p+t_0,t_p-t_0)|$ by varying $t_p$ and $t_0$. 

In section~\ref{sec:Kondo} we will demonstrate, for the example of the buildup of the Kondo peak, that this information is valuable to extract key features of the ongoing dynamics of the system, that are otherwise hidden by the uncertainty limit of standard probe pulses. The double probe measurement scheme may therefore open the path for the application of extremely short pulses to study the dynamics of emergent low-energy degrees of freedom. It can be noted that the coherence time of the features one wants to observe should be long enough to allow such two-probes measurement. However, the cases in which the double probe scheme is useful are exactly the ones in which the spectral features are sharp and, therefore, long-lived.

In the past, pioneering works~\cite{RevPetek1997} explored the possibilities of time-resolved {\em two}-photon photoemission, which uses two delayed pulses with energies ($\sim$3 eV) \textit{below} the work function of the sample to populate empty states and then photoemit electrons from there. The study of interferometric effects allowed to measure elastic and inelastic scattering rates of excited states at the surface of metals. It is important to underline the differences between this technique and the tomographic measurement proposed here. Because time-resolved two-photon photoemission allows only to study states with energies within 3 eV from the vacuum level, i.e states that are not occupied at equilibrium, it  probes a joint density of occupied and unoccupied states rather than the single-particle spectrum, and is therefore unsuited to study how the states below the Fermi energy change after a general kind of excitation, e.g. in an out-of-equilibrium phase transition or in the modification of the low energy properties of the system. In contrast, in the tomography described in this work the interference comes from different pathways leading to the same \textit{one}-photon photoemission event, which would allow to probe any bound state. 

\subsection{Photoemission with non-coherent states of light}
\label{sec:quantum}
Although the discussion in section~\ref{sec:Illustration} is based on the double pulse scheme, we now complete the generalization of the photoemission process extending it to 
states of the probe pulse which are not coherent states $|\alpha\rangle$, i.e. that have either statistical or quantum correlations. A coherent state $|\alpha\rangle$ is the closest quantum description of a classical wave and is therefore also called quasiclassical.
While the double probe experiment allows to reconstruct $G(t,t')$ over the full $(t,t')$ plane through a tomographic process, the manipulation of the 
state of the probe pulse 
beyond the coherent state-case
can allow to retrieve specific correlations in a single measurement.

Again, the discussion starts from $S(t,t')$. The probe pulse defines a filter $S(t,t')$ in the time-domain [Eq.~\eqref{eq:photoemission1}] or correspondingly in the time-frequency domain [Eq.~\eqref{eq:phototw}], which relates the signal $I(E,t_p)$ to the dynamical information contained in $G^<(t,t')$. In the analysis discussed so far the pulse autocorrelation function is generally limited to a form $S(t,t')=s(t)s(t')^*$ which can be {\em factorized} in the time-domain. On the other hand, an arbitrary function $S(t,t')$ could be designed to probe specific statistical features and correlations in the ongoing dynamics in the system.  To see how such a measurement might be implemented, we first note that an arbitrary function of Hermitian symmetry $S(t,t')=S(t',t)^*$ can be  expressed, through diagonalization, as the sum of factorizable functions:
\begin{equation}
\label{eq:s-eigen}
S(t,t') = \sum_j \, \eta_j \,\, \tilde s_j(t) \tilde s_j(t')^*,
\end{equation} 
where $j$ labels the  eigenvectors. If more than one $\eta_j$ is different from zero, then $S(t,t')$ is non-factorizable. From the above expression it can be seen that the result of the hypothetical photoemission with a non-factorizable filter $S(t,t')$ would be a weighted average $I = \sum_j \eta_j I_j$ of signals obtained with different pulses $\tilde s_j(t)$. This means that the measurement can be emulated by an equivalent ``tomographic'' set of experiments with factorizable filters. 

The usage of a factorizable $S(t,t')$ comes from the fact that in the standard case coherent states are considered for photoemission.
This opens the intriguing question whether the hypothetical tomographic experiment described in the previous paragraph can be replaced by a single measurement performed with 
non-coherent states of
light.  In order to see what kind of filters $S(t,t')$ one can obtain using 
general (quantum or statistical)
probe pulses, we will now show what differences arise if 
a general state is considered for the light in the photoemission process.

The derivation of the photoemission intensity for an arbitrary 
state (described by the density matrix $\rho$) of the incoming probe pulse is rather analogous to the 
semiclassical 
case discussed in Ref.~\cite{FreericksKrishnamurthyPruschke2009}, and it is therefore presented in the appendix. In the Coulomb gauge ($\bm \nabla \cdot \bm A =0$), the vector potential $\bm A$ is now given by an operator  $A(\bm r,t) = A^{+}(\bm r,t)  + A^{-}(\bm r,t)$
\cite{ScullyBook}, 
\begin{align}
\label{Aoperator}
A^{+}(\bm r,t) 
&
=
A^{-}(\bm r,t)^\dagger 
=
\sum_{{\bm q}}
\mathcal{A}_{\bm q}
\hat a_{\bm q} 
e^{i{\bm q}\bm r -i\omega_{\bm q}t}, 
\end{align}
where
$\mathcal{A}_{\bm q}=\sqrt{\hbar/ 2 V\epsilon_0 \omega_{\bm q}}$, and $\hat a_{\bm q}$ is the photon annihilation operator. For notational simplicity we take into account only one propagation direction $\hat n$ ($\bm q= q \hat n$) and transverse linear polarization. Setting the location of the sample at $\bm r=0$, the only change in Eq.~\eqref{eq:photoemission1} is that the autocorrelation function $S(t,t')$ takes its quantum-mechanical form  
\begin{align}
\label{eq:quantum-s}
S(t,t') = e^{i\Omega (t'-t)} \Tr \big(\rho\, A^{-}(t) A^{+}(t') \big).
\end{align}
In this context, classical coherent probe pulses can be described as a coherent state $|\Psi\rangle = |\aalpha\rangle = \prod_q |\alpha_q \rangle$, which immediately recovers the standard result, with $S_{\aalpha} (t,t') = s_{\aalpha}(t) s_{\aalpha}(t')^*$ and probe envelope $s_{\aalpha}(t) =  \sum_{q} e^{i (\omega_{q}-\Omega) t} \alpha_{q}^*\mathcal{A}_{q}$. On the other hand, if the state of the probe pulses is not a product $|\aalpha\rangle$ of coherent states, i.e. it is not a 
quasiclassical coherent wave,
$S(t,t')$ is not bound to the standard factorizable form. In fact, since the probe pulse only enters through the two-time correlation function, it can be shown that every function of the form \eqref{eq:s-eigen} can be obtained from Eq.~\eqref{eq:quantum-s} with a 
multimode state with Gaussian Wigner function
provided that $\eta_j\ge0$, i.e., the function is positive definite (see appendix \ref{sec:squeezed}).  In this way, a tomographic set of measurements aimed at reconstructing the effect of a Hermitian and positive definite filter $S(t,t')$ can, indeed, be replaced by a single experiment with 
probe pulses
not in a coherent state. 
In turn, every measurement with 
non-coherent states 
can be replaced by a tomographic set of measurements with classical pulses,  because any state of light $\rho$ can be expressed as an integral over coherent
states using the Glauber-Sudarshan P representation~\cite{1963Sudarshan,1963Glauber} $\rho = \int d\bbeta\, \mathcal{P}_\rho (\bbeta) |\bbeta\rangle\langle\bbeta|$.

The potential use of different filters $S(t,t')$ to obtain specific informations, such as the Wigner transform, is illustrated for the level quench in Fig.~\ref{fig:quantum}c, where we plot the result of Eq.~\eqref{eq:photoemission1} with a particular choice of an exotic, but hermitian and positive definite cross-shaped $S(t,t')$. As can be seen, the result shown in Fig.~\ref{fig:quantum}c resembles the Wigner transform plotted in Fig.~\ref{fig:quantum}b. In fact, an infinitely extending cross-shaped filter of the type used in  Fig.~\ref{fig:quantum}c would yield the sum of a constant background (from the diagonal $t=t'$) and the Wigner transform (from the antidiagonal $t=-t'+t_p$), from which the full Green's function can be reconstructed. On the actual experimental side, even though the generation and manipulation of quantum light pulses in the relevant spectral range for photoemission are not yet established, it is important to note that  corrections to the results of time-resolved photoemission can arise also if, instead of an enhancement of correlation, incoherent pulses are considered, which is also accounted for by the generalized expression. An incoherent light pulse has a reduced correlation between the various temporal positions in the pulse, as compared to coherent light. In the $(t,t')$ plane, $S(t,t')$ must quickly go to zero moving away from the $t=t'$ diagonal, as shown in the inset of Fig.~\ref{fig:quantum}d. This leads to a reduced frequency resolution [Fig.~\ref{fig:quantum}d] compared to the result obtained with a coherent pulse with the same intensity profile in time.

%%%%%%%%%%%%%%%%%%%%%%%%%%%%%%%%%%%%%%%%%%%%%%%%%%%%%%%%%%%%%%%%%
% KONDO SECTION
%%%%%%%%%%%%%%%%%%%%%%%%%%%%%%%%%%%%%%%%%%%%%%%%%%%%%%%%%%%%%%%%%

\section{Illustration and proposals for the double probe experiment}
\label{sec:Illustration}

\subsection{Buildup of the Kondo resonance: standard time-resolved photoemission}
\label{sec:Kondo}

In the following we illustrate the real-time measurement of the Green's function with the 
double-pulse technique (Fig.~\ref{fig:setup}) for the buildup of a Kondo resonance, which is a classical problem of nonequilibrium many-body physics and also a paradigmatic example of dynamics occurring beyond the spectral uncertainty limit. We start by discussing the results in standard photoemission.

The Kondo effect can arise when a localized orbital, such as a quantum dot or an impurity atom on a metallic surface, hybridizes with a continuum of conduction electrons \cite{HewsonBook}. Depending on the orbital occupancy, a magnetic moment is formed on the impurity when charge fluctuations are suppressed due to the Coulomb repulsion. This moment becomes screened by the conduction electrons below the Kondo temperature $T_K$, an emergent low energy scale of the system. The buildup of Kondo screening  in real time, e.g., after the impurity is suddenly coupled to the conduction band, has recently been a subject of intensive numerical research \cite{Nordlander1999, Anders2005, Antipov2015,Lechtenberg2015}. The spectroscopic signature of the Kondo effect is a narrow resonance of width $T_K$ at the Fermi energy (the so-called Kondo peak), which can be resolved only after times $\hbar/T_K$  \cite{Nordlander1999}, while the Kondo screening cloud is formed to a large extent on the much shorter timescale set by the hybridization between impurity and conduction band \cite{Lechtenberg2015}.  Thus the buildup of Kondo screening turns out to be a process which happens on timescales beyond the spectral uncertainty,  i.e. too fast to be spectrally resolved. 

\begin{figure}
\centering
\includegraphics[width=\columnwidth]{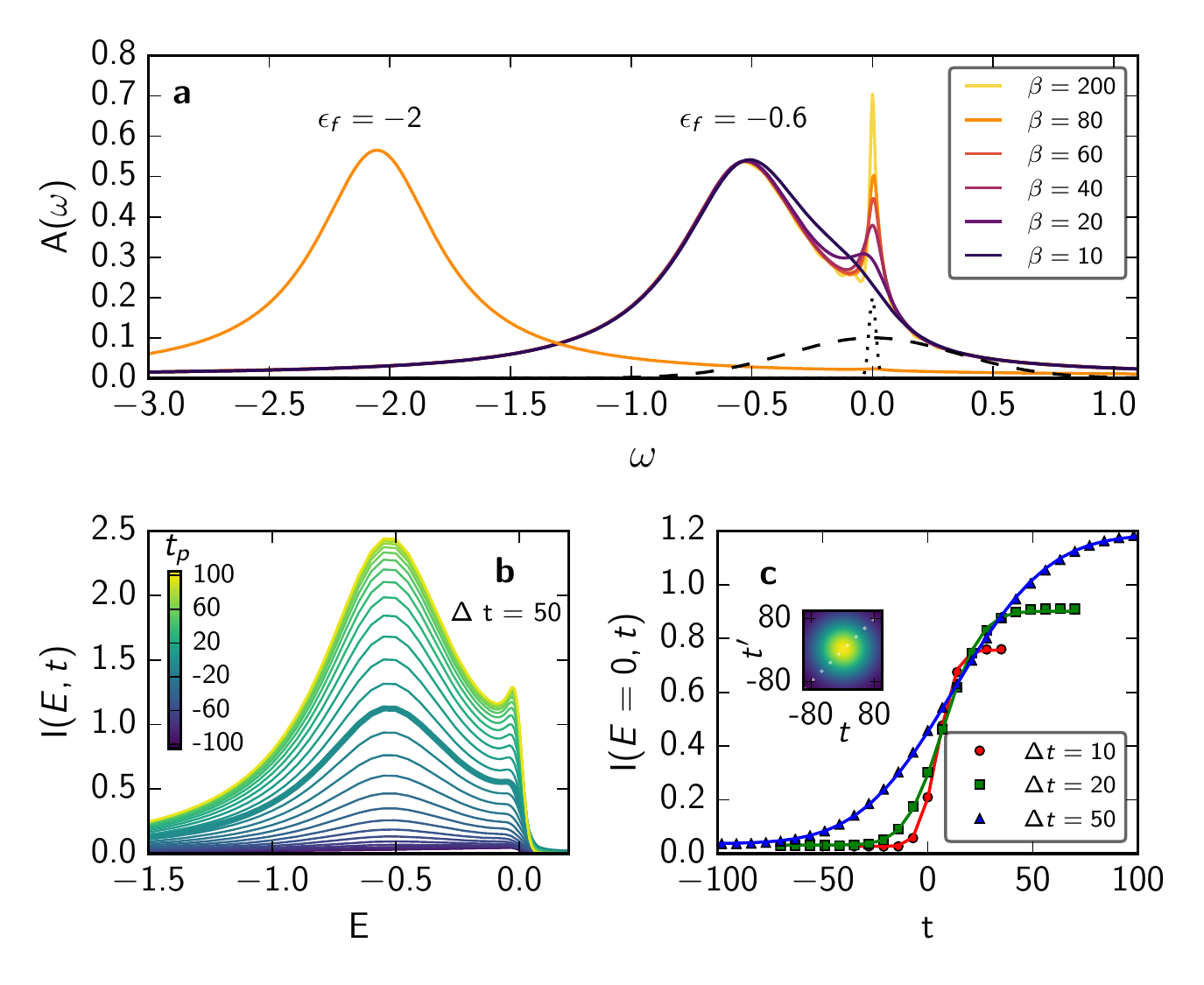}
\caption{
{\bf a)} Spectral function of the Kondo model  ($\Gamma_\text{dot}=0.6$) for various inverse temperatures $\beta$ at $\epsilon_f=-0.6$ ($T_K\approx 1/200$) and $\epsilon_f=-2$ ($T_K\approx 10^{-9}$). The dotted and dashed lines respectively show the power spectrum $\propto \exp(-\omega^2\Delta t^2)$ of Gaussian pulses with $\Delta t=50$ and $\Delta t=2$ for comparison. {\bf b)} Time-resolved photoemission spectrum $I(E,t_p)$ with a Gaussian probe of $\Delta t=50$. At $t=0$, $\epsilon_f$ is switched from $2$ to $-0.6$, the temperature is $T=1/80$. The probe time is varied in steps of $\Delta t_p=7$. The bold line corresponds to $t_p=0$. {\bf c)} Photo-emission intensity at $E=0$, obtained with different probe duration $\Delta t$. Lines are fits with an error function profile,  $a\,\text{erf}[(t-b)/\Delta t]+c$. Inset: $S(t,t')$ for $\Delta t = 50$.}
\label{fig:kondo_pes}
\end{figure}

We now discuss the Kondo effect at equilibrium and results of time-resolved photoemission during the buildup of the Kondo resonance within the Anderson model, 
\begin{multline}
H_{K}= \sum_{\sigma } \epsilon_f f_{\sigma}^\dagger f_{\sigma} + U f_{\uparrow}^\dagger f_{\uparrow} f_{\downarrow}^\dagger f_{\downarrow} + 
\\
+\sum_{k\sigma} V_{k } c_{k \sigma}^\dagger f_\sigma + h.c.  
+
\sum_{k\sigma}  \epsilon_{k} c_{k \sigma}^\dagger c_{k \sigma}.
\end{multline}
Here $f_\sigma$ and $c_{k\sigma}$ are annihilation operators for electrons with spin $\sigma$ on the impurity and bath levels, respectively, $\epsilon_f$ is the position of the impurity level relatively to the Fermi level,  $U$ the on-site Coulomb energy, and $V_k$ the tunneling matrix element between the impurity and the bath. We take the limit $U\to\infty$, so that a double occupancy of the level is suppressed, and assume a hybridization density of states $\Gamma(\epsilon) \equiv 2\pi\sum_{k} |V_k|^2 \delta(\omega-\epsilon_k) = \Gamma_\text{dot} / (e^{10(\epsilon/D-1)}+1)(e^{-10(\epsilon/D+1)}+1)$ which is constant $(\Gamma_\text{dot})$ below a smooth high-energy cutoff $D$. (In the following, $D/4$ and $4\hbar/D$ set the unit of energy and time, respectively.)  The Kondo temperature of the model depends exponentially on the position of the level  $T_K = D e^{-\pi |\epsilon_f|/\Gamma_\text{dot}}$. The Green's functions $G_f(t,t') = -i \langle  T_C f(t)  f^{\dagger} (t')\rangle $  at the impurity site is obtained within the time-dependent non-crossing approximation \cite{Nordlander1999}, using the implementation described in Ref.~\cite{Eckstein2010}. 

Figure \ref{fig:kondo_pes}a shows the spectral function $A_f(\omega)=-\frac{1}{\pi} \text{Im} G_f^{ret}(\omega+i0)$ of the impurity atom at equilibrium. The Lorentz peak of width $1/\Gamma_\text{dot}$ around $\omega\approx \epsilon_f$ represents the broadened impurity level. For $\epsilon_f=-0.6$ ($T_K\approx 1/200$), a Kondo resonance develops at the Fermi energy, whose width decreases proportionally to $T$ for $T \gtrsim T_K$. 
Following Ref.~\cite{Nordlander1999},  we now compute the buildup of the Kondo effect after the system is suddenly brought into the Kondo regime, by a shift of $\epsilon_f$ from a value $\epsilon_0=2$, for which the dot is basically empty, to the final value $\epsilon_1=-0.6$~\cite{Note2}. This may be thought of as a simplified representation of an experiment in which the orbital energy is suddenly shifted due to a core hole excitation. Figure  \ref{fig:kondo_pes}b shows the photoemission intensity from the impurity level  [Eq.~\eqref{eq:photoemission1}], obtained with a Gaussian probe $s(t)=\exp(-t^2/2\Delta t^2)$ with duration $\Delta t=50$, which is just long enough to resolve the final Kondo peak in frequency space (the bandwidth of the pulse is represented by the dotted line in Fig.~\ref{fig:kondo_pes}a). The buildup of spectral weight at $\omega=0$ can almost perfectly be fitted by an error function erf with rise time $\Delta t$ (c.f.~Fig.~\ref{fig:kondo_pes}c). This means that the buildup of the peak is resolution limited, consistently with Ref.~\cite{Nordlander1999}. With very short pulses $\Delta t < 1/\Gamma_\text{dot}$  the signal would no longer be temporally resolution limited (because $1/\Gamma_\text{dot} \approx 3.6$ sets the relaxation time of the impurity occupation, which is proportional to the total spectral weight), but such broadband pulses would completely wash out the Kondo peak.

\begin{figure}
\centering
\includegraphics[width=\columnwidth]{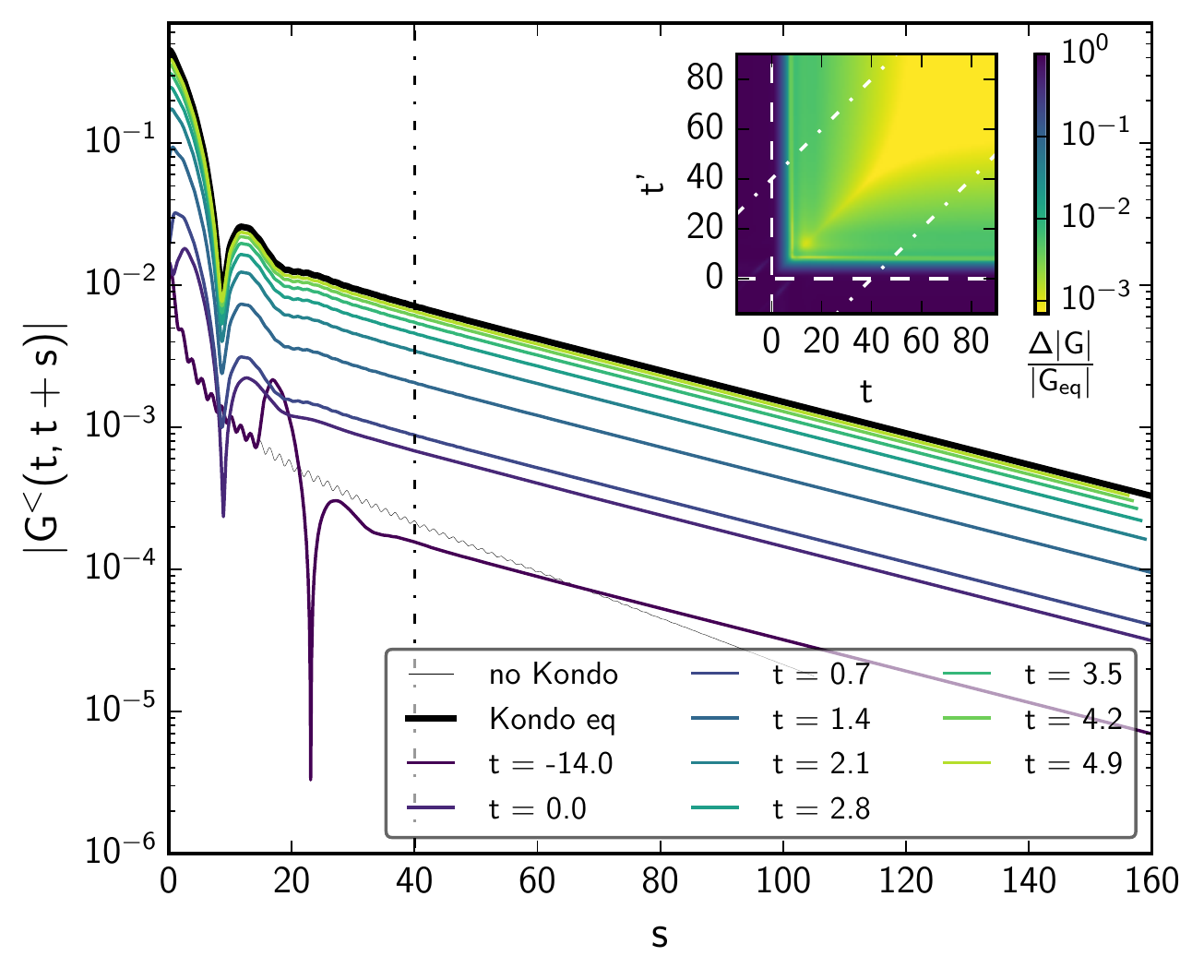}
\caption{Real-time Green's function $|G_f^<(t,t+s)|$ for the same parameters as in Fig.~\ref{fig:kondo_pes}b. The bold black line corresponds to the equilibrium result $|G^<_{eq}(t)|$ for $\epsilon_f=-0.6$ and $T=1/80$. Inset: Relative difference  $(|G_f^<(t,t')|- |G^<_{eq}(t-t')|)/ |G^<_{eq}(t-t')|$.}
\label{fig:kondo_realtime}
\end{figure}

\begin{figure}
\centering
\includegraphics[width=\columnwidth]{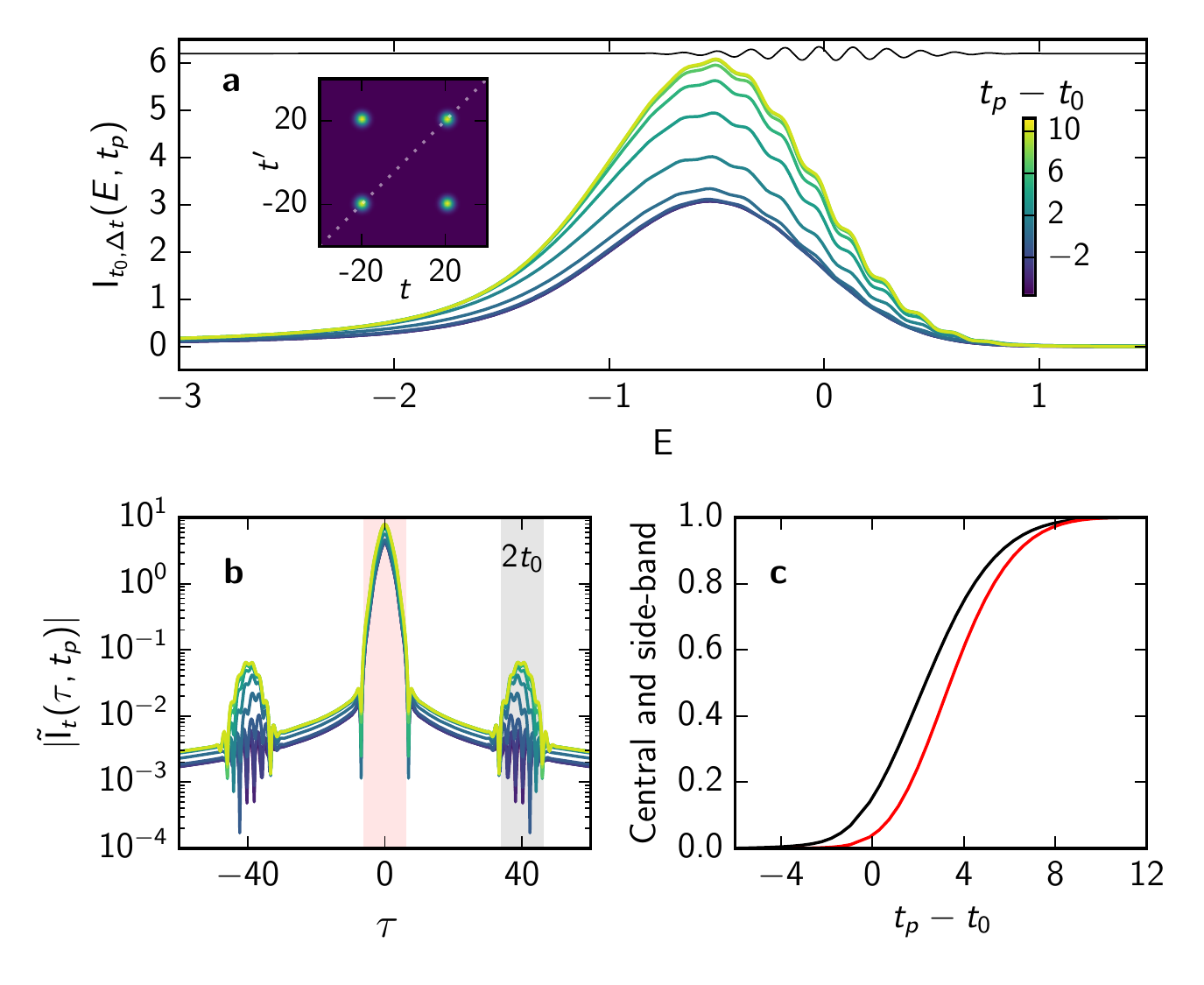}
\caption{
\textbf{a)} Photoemission signal $I(E,t_p)$ obtained with a double pulse [Eq.~\eqref{eq:double_pulse}] for the same parameters as in Fig.~\ref{fig:kondo_pes}b. The individual pulses are separated by $t_0=20$ and have a Gaussian envelope with duration $\Delta t=2$.
The black line shows the interference contribution [c.f., Eq.~\eqref{eq:photoemission-12}]. Inset: S(t,t') for the double pulse.
\textbf{b)} Fourier transform of the signal,  $\tilde I(\tau,t_p) = \int dE e^{iE\tau} I(E,t_p)$.
\textbf{c)} Fourier component of $\tilde I(\tau,t_p) = \int dE e^{iE\tau} I(E,t_p)$ at $\tau=2t_0$ (black line) and $\tau=0$ (red line), with normalized intensity.
}
\label{fig:pes_kondo_tomo}
\end{figure}

\subsection{Buildup of the Kondo resonance: double probe photoemission}
In spite of the uncertainty limited buildup of the Kondo peak, as measured via a standard photoemission experiment, an analysis of the Green's function in {\em real time} can reveal the underlying fast timescale. Figure \ref{fig:kondo_realtime} shows $|G_f^<(t,t+s)| = |\langle f^\dagger(t+s) f(t)\rangle |$ for fixed $t$ as a function of difference time $s$, corresponding to the hypothetical direct measurement of the decay of a hole on the impurity after it is created at time $t$. The fast initial drop at $s\lesssim 10$ corresponds to the Fourier transform of the bare energy level, while the slow exponential decay at large $s$ is related to the Kondo resonance. After the quench of the impurity level, we find that the equilibrium Kondo response is almost completely formed on a timescale $1/\Gamma_\text{dot}$, as soon as the population $n_f(t)=|G_f^<(t,t)|$ on the impurity is equilibrated. The same information is seen quantitatively from the inset in Fig.~\ref{fig:kondo_realtime}, which shows the relative difference between $|G^<_f(t,t')|$ and the equilibrated value reached for large times, $|G^<_{eq}(t-t')|$. 

The particular structure of the inset in Fig.~\ref{fig:kondo_realtime} can be seen as the real-time fingerprint that the relevant dynamics in the system occurs on timescales below the spectral uncertainty limit:  If $G^<(t,t')$ at given $s=t-t'$ equilibrates on a timescale $\tau_s \ll s$, this process cannot be resolved with a Gaussian probe pulse, because for sufficient time-resolution $\Delta t < \tau_s$ only the Green's function close to the $t=t'$ diagonal ($|t-t'| \lesssim \Delta t$) contributes to the photoemission signal  (see the inset of Fig.~\ref{fig:quantum}b). In contrast, the 
double-pulse technique (section~\ref{sec:tomography} and inset of Fig.~\ref{fig:setup}) is well suited to measure $|G^<(t,t')|$ at $t-t' \gg \Delta t$, i.e. well off from the diagonal. 

In order to illustrate the main steps to be performed, we plot in Fig.~\ref{fig:pes_kondo_tomo}a the result of a photoemission experiment with a double pulse [Eq.~\eqref{eq:double_pulse}] with time resolution $\Delta t=2$, and $t_0=20$, i.e. with pulses separated by $2t_0=40$. 
If we consider a Kondo peak of 20 meV in width, in physical units of time and energy, this setting would correspond to probe pulses of duration $\Delta t\simeq$ 12 fs and separated by $2t_0\simeq$ 240 fs.
As in the ideal case [Eq.~\eqref{eq:pestomo}] of ultra-short pulses, the signal is a superposition of an oscillating component $\propto \sin(2E t_0 + \varphi)$ due to the interference signal produced by the two
probe
pulses and a smooth background from the photoemission due to the individual pulses. For finite $\Delta t$ the contributions have a finite width $1/\Delta t$ in energy space~\cite{Note3} but they can nevertheless be separated by means of a Fourier transform of the signal with respect to $E$ (Fig.~\ref{fig:pes_kondo_tomo}b), provided that $\Delta t < t_0$, i.e. that the two probe pulses are separated in time. (Alternatively, the background may also be determined independently by averaging over the phase $\varphi$, which can be shifted by changing the relative carrier envelope phase of the two probes.)  This choice of $t_0$ allows to track the slow exponential decay of $|G^<(t,t+s)|$ (see dash-dotted lines in Fig.~\ref{fig:kondo_realtime}), which is related to the Kondo peak. To study its dynamics in this simple case, it is therefore not necessary to perform the full tomographic measurement, but it is sufficient just to vary $t_p$ at fixed $t_0$. In this way, the interference signal measures $|G(t_p+t_0,t_p-t_0)|$. The rapid increase of the amplitude of the oscillating signal around $t_p=t_0$ (Fig.~\ref{fig:pes_kondo_tomo}c) is an evidence of the characteristic fast relaxation dynamics in the Kondo problem (rise time of the black curve in Fig.~\ref{fig:pes_kondo_tomo}).
This demonstrates how the double probe scheme allows to analyze the relevant dynamics (the buildup of Kondo screening)  with arbitrary temporal resolution (Fig.~\ref{fig:pes_kondo_tomo}c), not limited by the inverse width of the Kondo peak.

\subsection{Melting of Mott gaps and amplitude mode in a superconductor}
The tomographic double probe scheme (Fig.~\ref{fig:setup}) could also be used to address other questions. How fast can a Mott gap be melted? What is the dynamics of the condensate of a superconductor when the Higgs mode is excited?
The true timescales of the destruction of electronic order in systems with charge density waves has not been clearly resolved as, for example, in the Mott-Peierls charge density wave $1T$-TaS$_2$ dichalcogenide. Recent experiments~\cite{Petersen2011,Sohrt2014} have determined that, while the lattice charge ordering is destroyed on the timescale of the relevant lattice mode, the electronic order is destroyed quasi-istantaneously on a resolution-limited timescale. Improving the temporal resolution in standard time-resolved photoemission experiments would unavoidably bring a worse spectral resolution, which is however important to resolve the dynamics of splitted bands. The double probe scheme would instead allow to access the true timescale of the process without giving up the spectral resolution.

Attention has also been dedicated to the excitation of the Higgs mode in BCS superconductors, such as Nb$_{1-x}$Ti$_{x}$N thin films.~\cite{Matsunaga2013,Matsunaga2014,Kemper2015} After the excitation with monocycle-like THz pulses, the transmittivity of the sample in the terahertz range oscillates at the frequency $2\Delta_{\mathrm{BCS}}$, as predicted for the excitation of the Higgs amplitude mode. Oscillations of the gap of a superconductor at its own frequency fall under the dynamics occurring at the spectral uncertainty limit. Photoemission with the tomographic double probe scheme is therefore a potential way to characterize the dynamics without the energy-time uncertainty limitation. As the lesser Green's function for laser-excited superconductors has been calculated theoretically, it would be interesting to evaluate the spectroscopic signature of the Higgs oscillations within the two-probe technique, and thus stimulate corresponding experiments.

%%%%%%%%%%%%%%%%%%%%%%%%%%%%%%%%%%%%%%%%%%%%%%%%%%%%%%%
%%%%%%%%%%%%%%%%%%%%%%%%%%%%%%%%%%%%%%%%%%%%%%%%%%%%%%%

\section{Conclusions}

In conclusion, we have analyzed possibilities to use time-resolved photoemission spectroscopy as a tool to probe the evolution of the electronic structure fully in the time-domain. Whenever relevant degrees of freedom evolve on timescales comparable to or faster than the inverse width of their spectral signatures, such a pure real-time characterization can provide more insight than a measurement in the  usual  mixed time-frequency domain, which requires a tradeoff between temporal and spectral resolution. 
In particular, we proposed 
a double probe pulse
technique, which can be used to probe the dynamics of low-energy degrees of freedom with arbitrarily short pulses, possibly in the attosecond range, with a bandwidth that would not be able to resolve the respective linewidths in frequency space if the pulses were taken alone. 

The case of dynamics occurring \textit{beyond the spectral uncertainty limit} is actually a quite common feature of correlated systems. As  examples, we discussed the buildup of the Kondo resonance, the melting of a Mott gap, and the dynamics of a superconductor when the amplitde mode is excited.
Further examples include the melting of other types of electronic order such as spin-density wave gaps~\cite{Tsuji2013} and the buildup of screening and plasmon resonances (in optics, the plasmon resonance at $\omega_p$ forms on a timescale $1/\omega_p$~\cite{Huber2001}).

Moreover, the use of quantum or statistical correlations in the light pulse would enhance the possibilities to characterize the dynamics with a time-resolved photoemission experiment. While this may seem technologically challenging at present, 
it contributes to the questions whether
correlations of light can be exploited to enhance, in a similar way, the capabilities of time-dependent measurements in other spectroscopic techniques, such as two-photon photoemission or optical spectroscopy.

\section*{Acknowledgments} 
We thank Fulvio Parmigiani for the insightful discussion and Sharareh Sayyad for the critical reading of the manuscript. F.R. was co-funded by the Friuli-Venezia Giulia region (European Social Fund, operative program 2007/2013). D.F. is financed by the European Research Council through the ERC Starting Grant n.677488, INCEPT.

%%%%%%%%%%%%%%%%%%%%%%%%%%%%%%%%%%%%%%%%%%%%%%%%%%%%
\appendix
%%%%%%%%%%%%%%%%%%%%%%%%%%%%%%%%%%%%%%%%%%%%%%%%%%%%

%%%%%%%%%%%%%%%%%%%%%%%%%%%%%%%%%%%%%%%%%%%%%%%%%%%%
\section{Theory of time-resolved photoemission spectroscopy with non-classical light pulses}
\label{appendix:derivation}
%%%%%%%%%%%%%%%%%%%%%%%%%%%%%%%%%%%%%%%%%%%%%%%%%%%%

To describe the photoemission process, we start from a general Hamiltonian
\begin{align}
H =
H_\text{matter}
+
H_\text{em}
+
H_\text{int},
\end{align}
where $H_\text{matter}$ describes both the solid and the outgoing electron states,
$H_\text{em}$ is the free Hamiltonian for the electromagnetic field, and $H_\text{int}$ the interaction
between matter and the radiation field. The interaction Hamiltonian for light and electrons with charge $-e$ is given by \cite{ScullyBook}
\begin{align}
H_\text{int} 
&=
H_\text{int}^{(1)} 
+
H_\text{int}^{(2)}
\\
H_\text{int}^{(1)} 
&=
\frac{e\hbar}{mc}\int d^3x\,
\hat \psi(\bm r)^\dagger
\big[
\hat{\bm A}(\bm r)
\cdot
{\frac{1}{i}\bm \nabla}
\big]
\hat \psi(\bm r)
\\
H_\text{int}^{(2)} 
&=
\frac{e^2}{2mc^2}
\int d^3x\,
\hat \psi(\bm r)^\dagger
\hat{\bm A}(\bm r)^2
\,
\hat \psi(\bm r),
\end{align}
where $\hat{ \bm A}(\bm r)$ is the operator for the vector potential, and $\hat \psi(\bm r)$ is the fermion field operator. Here and in the following we suppress the spin index. We expand the light-field in modes  with wave vector $\bm q$ and polarization 
$\lambda$, 
\begin{align}
H_\text{em}=
\sum_{\bm q,\lambda}
\hbar
\omega_{\bm q}\,
\hat a_{\bm q,\lambda}^\dagger
\hat a_{\bm q,\lambda},
\end{align}
so that the vector potential in Coulomb gauge ($\bm \nabla \cdot \bm A =0$) is given by
\begin{align}
\label{aplusaminus}
\hat{\bm A}(\bm r) 
&=
\hat{\bm A}^{+}(\bm r) 
+
\hat{\bm A}^{-}(\bm r),
\\
\label{Aoperator}
\hat{\bm A}^{+}(\bm r) 
&
=
\hat{\bm A}^{-}(\bm r)^\dagger 
=
\sum_{\bm q,\lambda}
\mathcal{A}_{\bm q}
\hat{\bm\epsilon}_{\bm q,\lambda}
\hat a_{\bm q,\lambda} e^{i\bm q \bm r -i\omega_{\bm q}t},
\end{align}
where $\hat{\bm\epsilon}_{\bm q,\lambda}$ are unit vectors with 
$\bm q \cdot \hat{\bm\epsilon}_{\bm q,\lambda}=0$, and 
\begin{align}
\mathcal{A}_{\bm q}=\sqrt{\frac{\hbar}{2 V\epsilon_0 \omega_{\bm q}}}.
\end{align}
Furthermore, we expand the matter field in a suitable basis
\begin{align}
\label{psiexpansion}
\hat \psi(\bm r)
=
\sum_{\bm k}
\phi_{\bm k}(\bm r)
\hat f_{\bm k}
+
\sum_{\alpha}
\chi_{\alpha}(\bm r)
\hat c_{\alpha},
\end{align}
where the index $\alpha$ refers to bound states (e.g., localized atomic wave functions or delocalized states in the solid), and $\bm k$ labels unbound states (outgoing waves) with asymptotic behavior $\phi_{\bm k}(\bm r) \sim e^{i\bm k \bm r}/\sqrt{V}$ and energy $E_{\bm k}=\hbar^2\bm k^2/2m+W$ (energies are considered with respect to the Fermi energy, and $W$ is the work function). The photoemission experiment measures the number of electrons that, under the effect of the light-matter interaction, are emitted into an initially unoccupied outgoing mode $\bm k$, i.e. the occupation probability $\langle\hat n_{\bm k}\rangle=\langle \hat f_{\bm k}^\dagger \hat f_{\bm k}\rangle$,
\begin{align}
\label{pp}
I_{\bm k} = \langle 
n_{\bm k}(t)
\rangle_{t\to\infty}
=
\big\langle
\mathcal{U}(t,t_0)^\dagger
\,n_{\bm k}\,
\mathcal{U}(t,t_0)
\big\rangle_0.
\end{align}
Here $\mathcal{U}(t,t')=T_{t} \exp [-i \int_{t'}^t d\bar t H(\bar t)]$ is the time-evolution operator, and $\langle  \cdots \rangle_0=\text{Tr}[ \rho_0 \cdots ] $ denotes the expectation value in the initial state $\rho_0$ for $t_0\to-\infty$, in which light and matter are uncorrelated and $ \langle \hat n_{\bm k} \rangle_0=0$ for all $\bm k$. 

The probability \eqref{pp} is computed in standard second-order time-dependent perturbation theory. We include all the non-perturbative processes that drive the system out-of-equilibrium (i.e. the pump pulse) in the time dependence of $H_{\mathrm{matter}}(t)$, and switch to the interaction representation with respect to $H_{\mathrm{matter}}(t)$, so that the time dependence of the operators is understood with respect to the uncoupled evolution $\mathcal{U}_0(t,t_0)=T_{t} \exp [-i \int_{t_0}^t d\bar t H_{\mathrm{matter}}(\bar t)]$. The full  time-evolution operator is expanded as 
\begin{equation}
\begin{split}
\mathcal{U}(t,t_0) &= \mathcal{U}_0(t,t_0) \biggl( 1 -i\int_{t_0}^t dt_1   H_\text{int}(t_1) \\
& -\int_{t_0}^t dt_1  \int_{t_0}^{t_1} dt_2 H_\text{int}(t_1) H_\text{int}(t_2) + \cdots\biggr).
\end{split}
\end{equation}
Because $f_{\bm k} $ gives zero when acting on $\rho_0=0$, the only non-vanishing contributions to Eq.~\eqref{pp} up to second order in the probe field are  
\begin{align}
\label{pp2}
I_{\bm k}
=
\lim_{t\to\infty}
\int_{t_0}^t dt_1  
dt_2
\,
\big\langle 
 H_\text{int}^{(1)}(t_1)
 \,
\hat n_{\bm k}(t)
 \,
 H_\text{int}^{(1)}(t_2)
\big\rangle_0.
\end{align}
To further simplify this expression, we rewrite $H_\text{int}^{(1)}$ using the expansion \eqref{psiexpansion}, 
\begin{align}
\label{hint11}
H_\text{int}^{(1)} 
=&
\frac{e\hbar i}{mc}
\sum_{\bm k,\alpha,j}
f_{\bm k}^\dagger
c_{\alpha}
\int\! d^3\bm r
A_j(\bm r)
\phi_{\bm k}(\bm r)^*
\nabla^j_{\bm r}
\chi_{\alpha}(\bm r)
+ h.c.,
\end{align}
where the sum over cartesian components $j$ is made explicit.
In this expression, we have kept only terms containing mixed products $f_{\bm k}^\dagger c_{\alpha} $ and $c_{\alpha}^\dagger f_{\bm k} $, which induce transitions between bound states and outgoing states. Terms proportional to $c_{\alpha'}^\dagger c_{\alpha} $ or $f_{\bm k'}^\dagger f_{\bm k}$ give no contribution in the expectation value \eqref{pp2}, as in this case an annihilation operator $f_{\bm k}$ would act on the initial state, which does not have any outgoing electron. 
To simplify the notation, we will assume linearly polarized light in the following, so that $A(\bm r, t)$ and $\nabla$ are understood as the components in the direction of the polarization. It is straightforward to reinsert the sums over cartesian components below. Inserting equation~\eqref{hint11} in equation~\eqref{pp2}, we thus have
\begin{multline}
\label{pp3}
I_{\bm k}
=
\int d1 d2
\,\biggl(\frac{e\hbar }{mc}\biggr)^2
\chi(1)^*\, \nabla_{1} \phi(1)
\,
\chi(2)\, \nabla_{2} \phi(2)^*
\,\times\,\\
\,\times\,\langle
\hat A(1)\,
\hat c(1)^\dagger\,
\hat f(1)\,
\hat n_{\bm k}(t)\,
\hat f(2)^\dagger\,
\hat c(2)\,
\hat A(2) \,
\rangle_0,
\end{multline}
with a combined notation for indices $1 \equiv (t_1,\bm r_1, \alpha_1, \bm k_1)$,
$\int d1 = \int_{t_0}^t dt_1 \int d^3\bm r_1 \sum_{\bm k_1,\alpha_1}$,
$\chi(1)=\chi_{\alpha_1}(\bm r_1)$,  $\phi(1)=\phi_{\bm k_1}(\bm r_1)$, 
$\hat A(1)=\hat A(\bm r_1,t_1)$, 
$\hat f(1)=\hat f_{\bm k_1}(t_1)$,
$\hat c(1)=\hat c_{\alpha_1}(t_1)$,
$\nabla_1=\nabla_{\bm r_1}$ (acting on $\bm r_1$).

The expectation value in the above integral can be factorized in a two-time correlation function of the field and a three-time correlation function of the electrons. To reduce the expression to the single particle properties of the solid alone, one commonly neglects the interaction of the outgoing electrons with the electrons within the solid. This so-called sudden approximation implies that the electronic correlation function factorizes in outgoing and bound state, so that the expectation value in \eqref{pp3} is given by the product of
\begin{align}
\langle
\hat f(1)
\hat n_{\bm k}(t)
\hat f(2)^\dagger
\rangle_0
=
e^{i(t_2-t_1)E_{\bm k}}
\delta_{\bm k_2,\bm k}
\delta_{\bm k_1,\bm k}\,\, ,
\end{align}
the lesser Green's function
\begin{align}
\langle
\hat c(1)^\dagger
\hat c(2)
\rangle_0
=
-iG^<(2,1)
=
-iG_{\alpha_2\alpha_1}^<(t_2,t_1),
\end{align}
and the light field correlation function 
\begin{align}
\label{psilight}
\Psi(1,2)
=
\langle
:\! \hat A(1) \hat A(2)\!:
\rangle_0.
\end{align}
Here $:\hat B:$ is the normal ordering, whose effect is to bring all annihilation operators to the right. The difference $\langle \hat A(1) \hat  A(2) \rangle_0 - \langle :\! \hat A(1) \hat A(2)\!: \rangle_0$ is the vacuum expectation value of $\langle \hat A(1) \hat A(2) \rangle_0$ (a pure number). Since we do not expect spontaneous photoemission from vacuum fluctuations, we can omit these terms.

As a further simplification, we assume that the propagation time $\delta t = L/c$ of light through the probed volume 
is small compared to the pulse duration $\Delta t$. Technically, the mode frequencies in the pulse are distributed around some large carrier frequency $\Omega$ and wave vector $\bm q_0$, with widths $\Delta \omega$ and $\Delta q = \Delta \omega /c$, respectively. We can then factor out this main carrier wave vector and set
\begin{align}
\label{eq:no-delay}
\hat{\bm A}^{\pm }(\bm r) \approx e^{\pm i \bm q_0 \bm r} \hat{\bm A}^{\pm}(\bm r=0),
\end{align} 
(with the probe volume centered at $\bm r=0$), provided that $L\Delta q \ll 1 $, which is indeed equivalent to $\delta t \ll \Delta t$ because $\Delta t \approx 1/\Delta \omega$. The approximation can be systematically improved, but, like any precise treatment of matrix elements, it would not alter the general discussion of the properties of time-resolved photoemission. Furthermore, the approximation is exact for a point-like object, like an atom on a surface, or a thin layer.  With these considerations and Eq.~\eqref{aplusaminus}, we can write the light correlation function as
\begin{align}
\Psi(1,2)
\approx
\sum_{\sigma_1,\sigma_2=\pm}
e^{i\sigma_1 \bm q_0 \bm r_1}
e^{i\sigma_2 \bm q_0 \bm r_2}
\Psi_{\sigma_1,\sigma_2}(t_1,t_2),
\end{align}
with
$
\Psi_{\sigma_1,\sigma_2}(t_1,t_2)=
\langle
\,:\!
\hat A^{\sigma_1}(t_1,0)
\hat A^{\sigma_2}(t_2,0)
\!:\,
\rangle_0$. Furthermore, it is useful to factor out the carrier frequency $\Omega$
\begin{align}
\Psi_{\sigma_1,\sigma_2}(t_1,t_2)
\equiv
e^{-i\Omega (\sigma_1 t_1 + \sigma_2 t_2)}
S_{\sigma_1,\sigma_2}(t_1,t_2).
\end{align}
The time-dependent part of the integral \eqref{pp3} is then written by
\begin{multline}
\sum_{\sigma_1,\sigma_2=\pm} (-i)\int_{t_0}^t dt_1  dt_2\, 
G_{\alpha_2\alpha_1}^<(t_2,t_1) S_{\sigma_1\sigma_2}(t_1,t_2)\,\times
\\
\,\times\,e^{i(t_2-t_1)E_{\bm k}} 
e^{i\Omega (\sigma_1 t_2 + \sigma_2 t_1)}.
\nonumber
\end{multline}
For large carrier frequency, only the term 
$(\sigma_1,\sigma_2)=(-,+)$ 
with exponential factor $e^{i(\Omega-E_{\bm k})(t_1-t_2)}$ will survive the  integration. For all the other combinations of $\sigma_1$ and $\sigma_2$,
the exponent contains 
counter-rotating  
terms which oscillate quickly compared to the time-dependence of $G$ and $S$, thus vanishing upon integration. The final result is
\begin{multline}
I_{\bm k} = 
\sum_{\alpha\alpha'}
p_{\bm q_0,\bm k,\alpha}^*
p_{\bm q_0,\bm k,\alpha'}
\,\,\,\times
\\
\times \,\,
\int dt  dt'
e^{i(t-t')(E_{\bm k}-\Omega)} 
(-i)G_{\alpha\alpha'}^<(t,t')
S(t,t'),
\label{pp5}
\end{multline}
where $p_{\bm q_0,\bm k,\alpha}$ are matrix elements between the bound and outgoing states
\begin{align}
p_{\bm q_0,\bm k,\alpha}
=
\int
\,
 d^3\bm r
 \,
e^{i\bm q_0 \bm r} \chi_{\alpha}(\bm r) \nabla \phi_{\bm k}(\bm r)^*,
\end{align}
and 
\begin{align}
S(t,t') 
&=
e^{i\Omega (t'-t)}
\sum_{q_1,q_2}
\mathcal{A}_{q_1}
\mathcal{A}_{q_2}
e^{i\omega_{q_1}t-i \omega_{q_2}t'}
\langle
a_{q_1}^\dagger
a_{q_2}
\rangle_0
\label{stt}
\nonumber\\
&
= e^{i\Omega (t'-t)}
\Tr \big (\rho\,  A^{-}(t) A^{+}(t') \big)
\end{align}
as in Eq.~\eqref{eq:quantum-s} of the main text. 

For 
a coherent state
the expectation value factorizes, so that 
$S(t,t')  =e^{-i\Omega t} \langle \Psi | A^{-}(t)| \Psi\rangle \langle \Psi | A^{+}(t') |\Psi\rangle e^{i\Omega t'} \equiv s(t)s(t')^*$ and $A(t) = \langle \Psi | A^{-}(t) | \Psi\rangle + \langle \Psi | A^{+}(t) |\Psi\rangle =s(t)e^{i\Omega t} + s(t)^*e^{-i\Omega t} $.

%%%%%%%%%%%%%%%%%%%%%%%%%%%%%%%%%%%%%%%%%%%%%%%%%%%%
\section{Pulse correlation for a multimode squeezed state}
\label{sec:squeezed}
%%%%%%%%%%%%%%%%%%%%%%%%%%%%%%%%%%%%%%%%%%%%%%%%%%%%

As an illustration of $S(t,t')$ for more general states with Gaussian Wigner function, 
we now evaluate Eq.~\eqref{stt} for a multi-mode squeezed vacuum
$|\Psi\rangle = Q^\dagger |0\rangle$, with the squeezing operator 
\begin{align}
Q=
\exp\Big[
\frac{1}{2}
\sum_{\omega \omega'}
R_{\omega,\omega'} a_{\omega}^\dagger a_{\omega'}^\dagger - h.c. \Big].
\label{squeeeeze}
\end{align}
The squeezing matrix $R$ is symmetric in frequency space, but not necessarily Hermitian.
(We switched between frequency and momentum labels $\omega = cq$ because light is propagating on one axis only.). 
We evaluate Eq.~\eqref{stt} in the frequency domain, i.e.,  we compute
\begin{align}
\label{sww}
S_{\omega,\omega'} 
&= \int dt dt' e^{-i\omega t }  S(t,t') e^{i\omega' t' }.
\end{align}
For simplicity, we set the energy shift $\Omega=0$ in the following. Inserting Eq.~\eqref{stt} in 
\eqref{sww} we obtain
\begin{align}
\label{sww2}
\frac{1}{(L/c)^2}
\mathcal{A}_{\omega} ^{-1}S_{\omega,\omega'}
\mathcal{A}_{\omega'} ^{-1}
&= \langle \Psi| a_{\omega}^\dagger a_{\omega'} |\Psi\rangle.
\end{align}
The prefactor is the normalization volume, with $\int dt e^{i(\omega-\omega')t} =(L/c)\delta_{\omega,\omega'}$. To evaluate the expectation value $\langle \Psi| a_{\omega}^\dagger a_{\omega'} |\Psi\rangle$,
we use the commutator relation $e^A B e^{-A} = B + [A,B] + \frac{1}{2!} [A,[A,B]]+...$ to obtain the general  identity \cite{ScullyBook},
\begin{align}
Q a_{\omega} Q^\dagger
=
\sum_{\omega'}
\mathcal{C}_{\omega,\omega'} a_{\omega'}
-
\sum_{\omega'}
\mathcal{S}_{\omega,\omega'} a_{\omega'}^\dagger,
\end{align}
with the functions
\begin{align}
\mathcal{C} = \sum_{n=0}^\infty \frac{(RR^\dagger)^n}{(2n)!} ,\,\,\,
\mathcal{S} =  \sum_{n=0}^\infty \frac{R(R^\dagger R)^n}{(2n+1)!}.
\label{sinh}
\end{align}
(For a single mode and $R = |r|e^{i\theta} \in \mathbb{C}$, the two functions correspond to the 
the hyperbolic functions $\mathcal{C} = \cosh{|r|}$ and $\mathcal{S} = e^{i\theta}\sinh{|r|}$.)
In the expectation value 
$\langle 0| Q a_{\omega}^\dagger a_{\omega'}  Q^\dagger |0\rangle$ we thus get, dropping the terms $a|0\rangle$,
\begin{align}
\langle 0| Q a_{\omega}^\dagger a_{\omega'}  Q^\dagger |0\rangle
&=
\langle 0| Q a_{\omega}^\dagger Q^\dagger Q a_{\omega'}  Q^\dagger |0\rangle
\\
&=
\sum_{\omega_1,\omega_2}
\mathcal{S}_{\omega,\omega_2}^*
\mathcal{S}_{\omega'\omega_1}
\langle 0|
 a_{\omega_2}
 a_{\omega_1}^\dagger
|0\rangle
\\
&=
(\mathcal{S}\mathcal{S}^\dagger)_{\omega',\omega},
\label{sww1}
\end{align}
which together with Eq.~\eqref{sww2} concludes the determination of $S_{\omega,\omega'}$.

More interestingly, we can use Eqs.~\eqref{sww2} and \eqref{sww1} to prove the statement, given in the main text, that any desired correlation function $S_{\omega,\omega'}$  (or $S(t,t')$) can be, in principle, obtained from a {\em single} light pulse, provided that the matrix $S_{\omega,\omega'}$ is (i), {\em Hermitian},  and (ii), {\em positve definite}. Conditions (i) and (ii) imply that the matrix $M_{\omega,\omega'}\equiv\frac{1}{(L/c)^2}\mathcal{A}_{\omega} ^{-1}S_{\omega,\omega'}\mathcal{A}_{\omega'} ^{-1}$  in Eq.~\eqref{sww2} can be diagonalized, $M = V d V^\dagger$, where $d$ is diagonal with $d_{\alpha\alpha}\ge 0$. The latter implies that the choice
\begin{align}
R = V \text{asinh}(\sqrt{d}) V^*,
\label{the_r}
\end{align}
with $(V^*)_{\omega,\omega'}=V_{\omega,\omega'}^*$, is a well-defined symmetric matrix. Using Eqs.~\eqref{sinh} and \eqref{sww1} one can then directly verify that Eq.~\eqref{sww2} is satisfied, i.e, the squeezed vacuum Eq.~\eqref{squeeeeze} with the squeezing matrix \eqref{the_r} gives the desired cross-correlation $S_{\omega,\omega'}$.


\begin{thebibliography}{100}
\newcommand{\bibtitle}[1]{``#1''}

\bibitem{Damascelli2003}
A.~Damascelli, Z.~Hussain, and Z.~Shen,
\bibtitle{Angle-resolved photoemission studies of the cuprate superconductors},
Rev. Mod. Phys. {\bf 75}, 473 (2003).

\bibitem{Avella}
\bibtitle{Strongly Correlated Systems: Experimental Techniques}, 
edited by A.~Avella and F.~Mancini, 
Springer Series in Solid-State Sciences, Springer Verlag Berlin Heidelberg.

\bibitem{Giannetti2016}
C.~Giannetti, M.~Capone, D.~Fausti, M.~Fabrizio, F.~Parmigiani, and  D.~Mihailovic,
\bibtitle{Ultrafast optical spectroscopy of strongly correlated materials and high-temperature superconductors: a non-equilibrium approach},
arXiv:1601.07204.

\bibitem{Schmitt2008}
F.~Schmitt et~al.
\bibtitle{Transient electronic structure and melting of a charge density wave in TbTe$_3$},
Science {\bf 321} 1649, (2008).

\bibitem{Rohwer2011}
T.~Rohwer et~al.
\bibtitle{Collapse of long-range charge order tracked by time-resolved photoemission at high momenta},
Nature {\bf 471}, 490 (2011).

\bibitem{Petersen2011}
J.C. Petersen, S. Kaiser, N. Dean, A. Simoncig, H.Y. Liu, A.L. Cavalieri, C. Cacho, I.C.E. Turcu, E. Springate, F. Frassetto, L. Poletto, S.S. Dhesi, H. Berger, and A. Cavalleri
\bibtitle{Clocking the melting transition of charge and lattice order in 1$T$-TaS$_2$ with ultrafast extreme-ultraviolet angle-resolved photoemission spectroscopy}
Phys. Rev. Lett. {\bf 107}, 177402 (2011).

\bibitem{Perfetti2006}
L.~Perfetti, P.~A. Loukakos, M.~Lisowski, U.~Bovensiepen, H.~Berger,  S.~Biermann, P.~S. Cornaglia, A.~Georges, and M.~Wolf,
\bibtitle{Time evolution of the electronic structure of $1T\mathrm{\text{-}}{\mathrm{TaS}}_{2}$ through the insulator-metal transition},
Phys. Rev. Lett. {\bf 97}, 067402 (2006).

\bibitem{Wegkamp2014}
D.~Wegkamp, M.~Herzog, L.~Xian, M.~Gatti, P.~Cudazzo,  Ch.~L. McGahan, R.~E. Marvel, R.~F. Haglund, A.~Rubio,
  M.~Wolf, and J.~St\"{a}hler,
  \bibtitle{Instantaneous band gap collapse in photoexcited monoclinic  ${\mathrm{VO}}_{2}$ due to photocarrier doping},
Phys. Rev. Lett. {\bf 113}, 216401 (2014).

\bibitem{Chini2014}
M.~Chini, K.~Zhao, and Z.~Chang.
\bibtitle{The generation, characterization and applications of broadband isolated attosecond pulses},
Nat. Phot., {\bf 8}, 178 (2014).

\bibitem{Lechtenberg2015}
B.~Lechtenberg and F.~B. Anders,
\bibtitle{Spatial and temporal propagation of Kondo correlations},
Phys. Rev. B {\bf 90}, 045117 (2014).

\bibitem{Matsunaga2013}
R. Matsunaga, Y.I. Hamada, K. Makise, Y. Uzawa, H. Terai, Z. Wang, and R. Shimano
\bibtitle{Higgs amplitude mode in the BCS superconductor Nb$_{1-x}$Ti$_x$N induced by terahertz pulse excitation}
Phys. Rev. Lett. {\bf 111}, 057002 (2013).

\bibitem{Matsunaga2014}
R. Matsunaga, N. Tsuji, H. Fujita, A. Sugioka, K. Makise, Y. Uzawa, H. Terai, Z. Wang, H. Aoki, and R. Shimano
\bibtitle{Light-induced collective pseudospin precession resonating with Higgs mode in a superconductor}
Science {\bf 345}, 1145 (2014).

\bibitem{MukamelBook}
S. Mukamel
\bibtitle{Principles of Nonlinear Optics and Spectroscopy}
Oxford University Press, Oxford (1995).

\bibitem{Mukamel1999}
W. M. Zhang, V. Chernyak, and S. Mukamel
\bibtitle{Multidimensional femtosecond correlation spectroscopies of electronic and vibrational excitons}
J. Chem. Phys. {\bf 110}, 5011 (1999).

\bibitem{Sohrt2014}
C. Sohrt, A. Stange, M. Bauer, and K. Rossnagel
\bibtitle{How fast can a Peierls-Mott insulator be melted?}
Faraday Discuss. {\bf 171}, 243 (2014).

\bibitem{Kemper2015}
A.~F. Kemper, M.~A. Sentef, B.~Moritz, J.~K. Freericks, and T.~P. Devereaux,
\bibtitle{Direct observation of Higgs mode oscillations in the pump-probe photoemission spectra of electron-phonon mediated superconductors},
Phys. Rev. B {\bf 92}, 224517 (2015).

\bibitem{Haroche2012}
S~Haroche, 
\bibtitle{Nobel lecture: Controlling photons in a box and exploring the quantum to classical boundary},
Rev. Mod. Phys. {\bf 85}, 1083 (2013).

\bibitem{Wineland2012}
D.~Wineland,
\bibtitle{Nobel lecture: Superposition, entanglement, and raising  Schr\"odinger's cat},
Rev. Mod. Phys., {\bf 85}, 1103 (2013).

\bibitem{Schnabel2010}
R.~Schnabel, N.~Mavalvala, D.~E. McClelland, and P.~K. Lam,
\bibtitle{Quantum metrology for gravitational wave astronomy},
Nat. Comm. {\bf 1}, 121 (2010).

\bibitem{Treps2002}
N.~Treps, U.~Andersen, B.~Buchler, P.~K. Lam, A.~Ma\^{\i}tre, H.-A. Bachor, and C.~Fabre,
\bibtitle{Surpassing the standard quantum limit for optical imaging using nonclassical multimode light},
Phys. Rev. Lett., {\bf 88}, 203601 (2002).

\bibitem{Esposito2015}
M.~Esposito, K.~Titimbo, K.~Zimmermann, F.~Giusti, F.~Randi, D.~Boschetto,  F.~Parmigiani, R.~Floreanini, F.~Benatti, and D.~Fausti,
\bibtitle{Photon number statistics uncover the flucutuations in non-equilibrium lattice dynamics},
Nat. Comm. {\bf 6}, 10249 (2015).

\bibitem{FreericksKrishnamurthyPruschke2009}
J.~K. Freericks, H.~R. Krishnamurthy, and Th. Pruschke.
\bibtitle{Theoretical description of time-resolved photoemission spectroscopy: Application to pump-probe experiments},
Phys. Rev. Lett., {\bf 102}, 136401 (2009).

\bibitem{Eckstein2008c}
M.~Eckstein and M.~Kollar,
\bibtitle{Measuring correlated electron dynamics with time-resolved photoemission spectroscopy},
Phys. Rev. B, {\bf 78}, 245113 (2008).

\bibitem{Note0}
Note that this formulation adopts a one-step model of photoemission in which the interaction of the outgoing electrons and the electrons within the solid is neglected (sudden approximation), a common approximation beyond which photoemission cannot be expressed in terms of the properties of the solid alone.

\bibitem{Note1}
There is an analogy of Eq.(\ref{eq:phototw}) to the case of a Wigner quasi-probability 
distribution in phase space, which becomes strictly non-negative only when averaged 
over phase space regions of size $\Delta x \Delta p  \gg \hbar$.

\bibitem{Eberly1977}
J.~Eberly and K.~Wòdkiewicz
\bibtitle{The time-dependent physical spectrum of ligh}
J. Opt. Soc. Am. 67 {\bf 9}, 1252 (1997)

\bibitem{Mukamel1996}
S.~Mukamel, C.~Ciordas-Ciurdariu and V.~Khidekel
\bibtitle{Wigner Spectrograms for Femtosecond Pulse-Shaped Heterodyne and Autocorrelation Measurements}
IEEE J. Quantum Electron. 32 {\bf 8}, 1278 (1996)

\bibitem{Dorfman2012}
K.~Dorfman and S.~Mukamel
\bibtitle{Nonlinear spectroscopy with time- and frequency-gated photon counting: A superoperator diagrammatic approach}
Phys. Rev. A {\bf 86}, 013810 (2012)

\bibitem{RevPetek1997}
H.~Petek and S.~Ogawa,
\bibtitle{Femtosecond time-resolved two photon photoemission studies of electron dynamics in metals},
Progress in Surface Science, {\bf 56}, 293 (1997).

\bibitem{ScullyBook}
O.~M. Scully and M.~S. Zubairy,
\bibtitle{Quantum Optics},
Cambridge University Press, Cambridge (1997).

\bibitem{1963Glauber}
R.~J. Glauber.
\bibtitle{Coherent and incoherent states of the radiation field},
Phys. Rev. {\bf 131}, 2766 (1963).

\bibitem{1963Sudarshan}
E.~C.~G. Sudarshan.
\bibtitle{Equivalence of semiclassical and quantum mechanical descriptions of statistical light beams},
Phys. Rev. Lett., {\bf 10}, 277 (1963).

\bibitem{HewsonBook}
A.~C. Hewson,
\bibtitle{The Kondo problem to heavy fermions},
Cambridge University Press, Cambridge (1993).

\bibitem{Nordlander1999}
P.~Nordlander, M.~Pustilnik, Y.~Meir, N.~S.~Wingreen, and D.~C.~Langreth,
\bibtitle{How long does it take for the Kondo effect to develop?},
Phys. Rev. Lett. {\bf 83}, 808 (1999).

\bibitem{Anders2005}
F.~B. Anders and A.~Schiller.
\bibtitle{Real-time dynamics in quantum-impurity systems: A time-dependent numerical renormalization-group approach},
Phys. Rev. Lett. {\bf 95}, 196801 (2005).

\bibitem{Antipov2015}
A.~Antipov, Q.~Dong, and E.~Gull.
\bibtitle{Voltage quench dynamics of a Kondo system},
arXiv:1508.06633.

\bibitem{Eckstein2010}
M.~Eckstein and Ph.~Werner.
\bibtitle{Nonequilibrium dynamical mean-field calculations based on the noncrossing approximation and its generalizations},
Phys. Rev. B, {\bf 82}, 115115 (2010).

\bibitem{Note2}
The precise ramp profile is $\epsilon_f(t) = (\epsilon_1-\epsilon_0) [1-\cos(2\pi t/t_c)]/2 + \epsilon_0$ within a short ramp time $0<t<t_c=1$.

\bibitem{Note3}
Note that the two components are centred around the location $\epsilon _f$ of the bare energy level and the Kondo peak, respectively.

\bibitem{Tsuji2013}
N.~Tsuji, M.~Eckstein, and Ph.~Werner, 
\bibtitle{Nonthermal antiferromagnetic order and nonequilibrium criticality in the Hubbard model},
Phys. Rev. Lett. {\bf 110}, 136404 (2013).

\bibitem{Huber2001}
R.~Huber, F.~Tauser, A.~Brodschelm, M.~Bichler, G.~Abstreiter, and  A.~Leitenstorfer,
\bibtitle{How many-particle interactions develop after ultrafast excitation of an electron-hole plasma},
Nature {\bf 414}, 286 (2011).


\end{thebibliography}
\end{document}